\documentclass{IEEElmag}

\usepackage[colorlinks,urlcolor=blue,linkcolor=blue,citecolor=blue]{hyperref}

\usepackage[hyphenbreaks]{breakurl}
\usepackage{color,array}
\usepackage{graphicx}
\usepackage{CJKutf8}
\usepackage{tabulary}
\usepackage{colortbl}

\usepackage{cite}
\usepackage[symbol]{footmisc}

\jvol{XX}
\jnum{XX}
\paper{1234567}
\pubyear{2024}
\receiveddate{xxxx 00, 0000}
\reviseddate{xxxx 00, 0000}
\publisheddate{xxxx 00, 0000}
\currentdate{xxxx 00, 0000  (Dates will be inserted by IEEE; ``published'' is the date the accepted preprint is posted on IEEE Xplore\textsuperscript{\textregistered}; ``current version'' is the date the typeset version is posted on Xplore\textsuperscript{\textregistered})}
\doiinfo{LMAG.2020.Doi Number}

\begin{document}

\sptitle{Spin Electronics}

\title{One Trillion True Random Bits Generated with a Field Programmable Gate Array Actuated Magnetic Tunnel Junction}

\author{Andre Dubovskiy\affilmark{1}} 
\author{Troy Criss\affilmark{1}}
\author{Ahmed Sidi El Valli\affilmark{1}}
\author{{Laura Rehm}\affilmark{1}}
\author{Andrew D. Kent\affilmark{1}*}
\author{Andrew~Haas\affilmark{1}}

\affil{Department of Physics, New York University, New York, NY 10003 USA}

\IEEEmember{*Fellow, IEEE}
\corresp{Corresponding authors: Andre Dubovskiy (\href{mailto:dubovski@nyu.edu}{\color{blue}dubovski@nyu.edu}), Andrew D. Kent (\href{mailto:andy.kent@nyu.edu}{\color{blue} andy.kent@nyu.edu}) and Andrew Haas (\href{mailto:andy.haas@nyu.edu}{\color{blue} andy.haas@nyu.edu})}

\markboth{Preparation of Papers for \emph{IEEE Magnetics Letters}}{Author Name}

\begin{abstract}
Large quantities of random numbers are crucial in a wide range of applications. We have recently demonstrated that perpendicular nanopillar magnetic tunnel junctions (pMTJs) can produce true random bits when actuated with short pulses. However, our implementation used high-end and expensive electronics, such as a high bandwidth arbitrary waveform generator and analog-to-digital converter, and was limited to relatively low data rates. Here, we significantly increase the speed of true random number generation (TRNG) of our stochastic actuated pMTJs (SMART-pMTJs) using Field Programmable Gate Arrays (FPGAs), demonstrating the generation of over $10^{12}$ bits at rates exceeding 10Mb/s. The resulting bitstreams pass the NIST Statistical Test Suite for randomness with only one XOR operation. In addition to a hundred-fold reduction in the setup cost and a thousand-fold increase in bitrate, the advancement includes simplifying and optimizing random bit generation with a custom-designed analog daughter board to interface an FPGA and SMART-pMTJ. The resulting setup further enables FPGA at-speed processing of MTJ data for stochastic modeling and cryptography.
\end{abstract}


\maketitle

\section{INTRODUCTION}\label{Sec:Intro}
Probabilistic computing architectures have been proposed to efficiently solve real-world problems with uncertain or complex input data {[}Feynman 1982{]}. Applications include cryptography {[}McInnes 1991{]}, neuromorphic systems {[}Schuman 2017{]}, and Monte Carlo simulations {[}Harrison 2010{]}. The acquisition of probabilistic bits, or p-bits, needed for these uses can be a technical challenge of its own. In fact, certain particle physics collision simulations spend more than half of the computational time simply generating random numbers {[}Misra 2023{]}. These random numbers are also often pseudo-random, produced in software from a seed with chaotic but deterministic algorithms.

True random number generators (TRNGs) for use in probabilistic applications promise a significant improvement in operational speed and efficiency. Existing CMOS solutions typically require large numbers of transistors~{[}Yang 2014; Nguyen 2020{]} or have practical limitations in terms of size, speed, and operating conditions~{[}Zhun 2001; Herrero-Collantes 2017{]}. Thus the development of TRNG devices that are compact, fast, energy-efficient, and reliable is an important goal. 

Magnetic noise offers a promising source of true random numbers. This is because nanometer scale ferromagnetic elements can function as two-state systems, with their magnetization states ``up'' or ``down,'' separated by an energy barrier denoted as $E_{b}$. When the energy barrier is on a similar scale as the thermal energy $kT$ --- where $k$ is the Boltzmann constant and $T$ is the operating temperature of the device --- the magnetization fluctuates between these two states, a phenomenon referred to as superparamagnetism.

\begin{figure}[t]
\centering\includegraphics[width=0.8\linewidth]{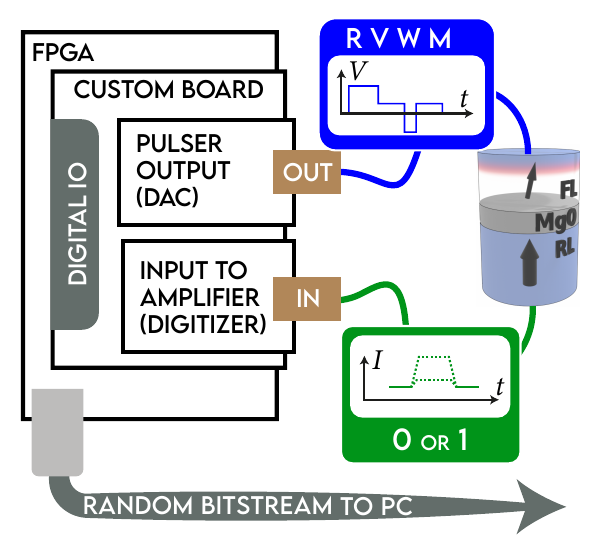}
\caption{A custom-designed circuit interfaces an FPGA to a pMTJ. A DAC is used to apply a reset (R)-verify (V)-write (W)-measure (M) pulse sequence, and a transimpedance amplifier converts the current flow through the MTJ into a voltage input. The current flow during the measurement pulse is used to create a sequence of 0s and 1s that are queued and sent to a computer. A schematic of the pMTJ is shown on the right. It consists of a reference layer (RL), a tunnel barrier (MgO), and a free layer (FL). The orientation of the free layer's magnetic moment can be parallel (P) or anti-parallel (AP) to that of the reference layer, resulting in low and high resistance states, respectively.}
\label{Fig:Fig1}
\end{figure}

\begin{figure}[h]
\centering\includegraphics[width=0.8\linewidth]{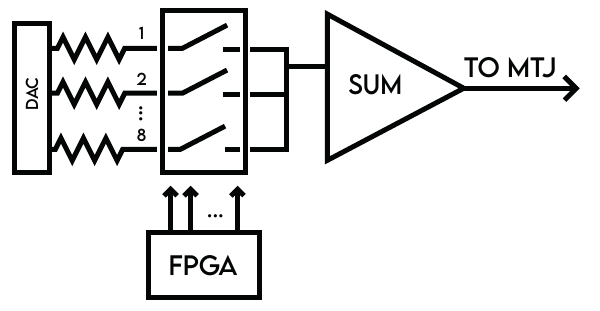}
\caption{The FPGA controls voltages from the DACs via analog switches based on the pulse timings. These voltages are added and output to the junction.}
\label{Fig:Fig2}
\end{figure}

A device known as a magnetic tunnel junction (MTJ) can transform the magnetization fluctuations into easily detectable two-level electrical signals. Additionally, MTJs can be seamlessly incorporated with complementary metal-oxide semiconductor (CMOS) technology, as highlighted in references~{[}Prenat 2007; Matsunaga 2008; Zhao 2008; Kent2015; Deng 2016; Kumar 2019; Barla 2020; Barla 2020{]}. Notably, there has been an extensive exploration into the use of MTJs with a superparamagnetic layer for the purpose of probabilistic computing, as detailed in references ~{[}Vodenicarevic 2017; Borders 2019; Kaiser 2019; Parks 2018; Hayakawa 2021; Safranski 2021{]}.

In contrast to superparamagnetic MTJs, Rehm~\emph{et al.} showed that stable perpendicularly magnetized MTJs (pMTJs) ($E_{b} \approx 39 kT$) actuated by pulses to be an excellent alternative approach to TRNG {[}Rehm 2023{]}. Using a stochastic write pulse with a switching probability $P$ of close to 50\% to generate each bit results in a stream of true random bits, much like a series of coin flips results in a stream of heads or tails. The generated bit stream can be much less sensitive to device operating temperature, device-to-device variations, and other operating conditions~{[}Rehm 2023; Morshed 2023{]}; the device was denoted a stochastic magnetic actuated random transducer, or SMART device.

The key advancement we will describe here is the simplification, optimization, and dramatic acceleration of the random bit generation from a SMART device with the help of a field programmable gate array (FPGA) and a custom-designed analog daughter board to interface the FPGA
with the SMART-pMTJ.

\section{EXPERIMENTAL SETUP}
To generate random bits with MTJs, we used a custom FPGA daughter board and associated FPGA (Intel Cyclone 10 GX FPGA, qDK-DEV-10CX220-A) as illustrated schematically in Fig.~\ref{Fig:Fig1}. First, we create and send voltage pulses across the junction. The pulses are created from a custom voltage source by toggling analog switches (part \# SN74AUC2G66DCTR) controlled by the FPGA with a cyclic state machine in Verilog. Since the FPGA itself can only output boolean signals, we made an analog daughter board that creates constant voltage sources with an 8-channel, 12-bit digital-to-analog converter (part \# DAC7578SPWR) to adjust the amplitudes of the pulses. Four voltages are inverted with an additional amplifier stage to create negative outputs. All eight outputs' adjusted voltages are then added with an operational amplifier and combined into one output that gets sent to the MTJ, as shown in Fig.~\ref{Fig:Fig2}.

While outputting the voltage pulses, we measure the junction current. A transimpedance amplifier (part \# OPA699ID) on the daughter board converts the current flowing through the junction to a voltage that the FPGA can measure. This circuit virtually grounds one terminal of the MTJ and transforms a $\simeq 20\mu$A current to the order of 2V. This allows the FPGA to determine if the junction is in a high or low resistance state, recorded as bit 0 or 1. The high and low resistance states correspond to approximately 2k$\Omega$ and 1k$\Omega$. The probabilistic bit is the result of a write pulse that switches the junction state approximately 50\% of the time, i.e., a stochastic write. A measure pulse is used to determine if the junction switched to a parallel (P) state, corresponds to a bit 1, or the junction remained in an anti-parallel (AP) state, a bit 0. The amplified signal is sampled by the FPGA at a rate of 468.75MHz. The firmware of the FPGA selects one sample per pulse cycle of the experiment during the measure pulse to collect one random bit per pulse cycle.  Finally, the random bits from the FPGA are sent to a computer over a 10Gb/s Ethernet User Datagram Protocol (UDP) connection to record bit stream and for further data analysis. 

\section{MEASUREMENT PROCEDURE}
\begin{figure}[h]
    \centering\hspace*{0.71cm}\includegraphics[width=0.9\linewidth]{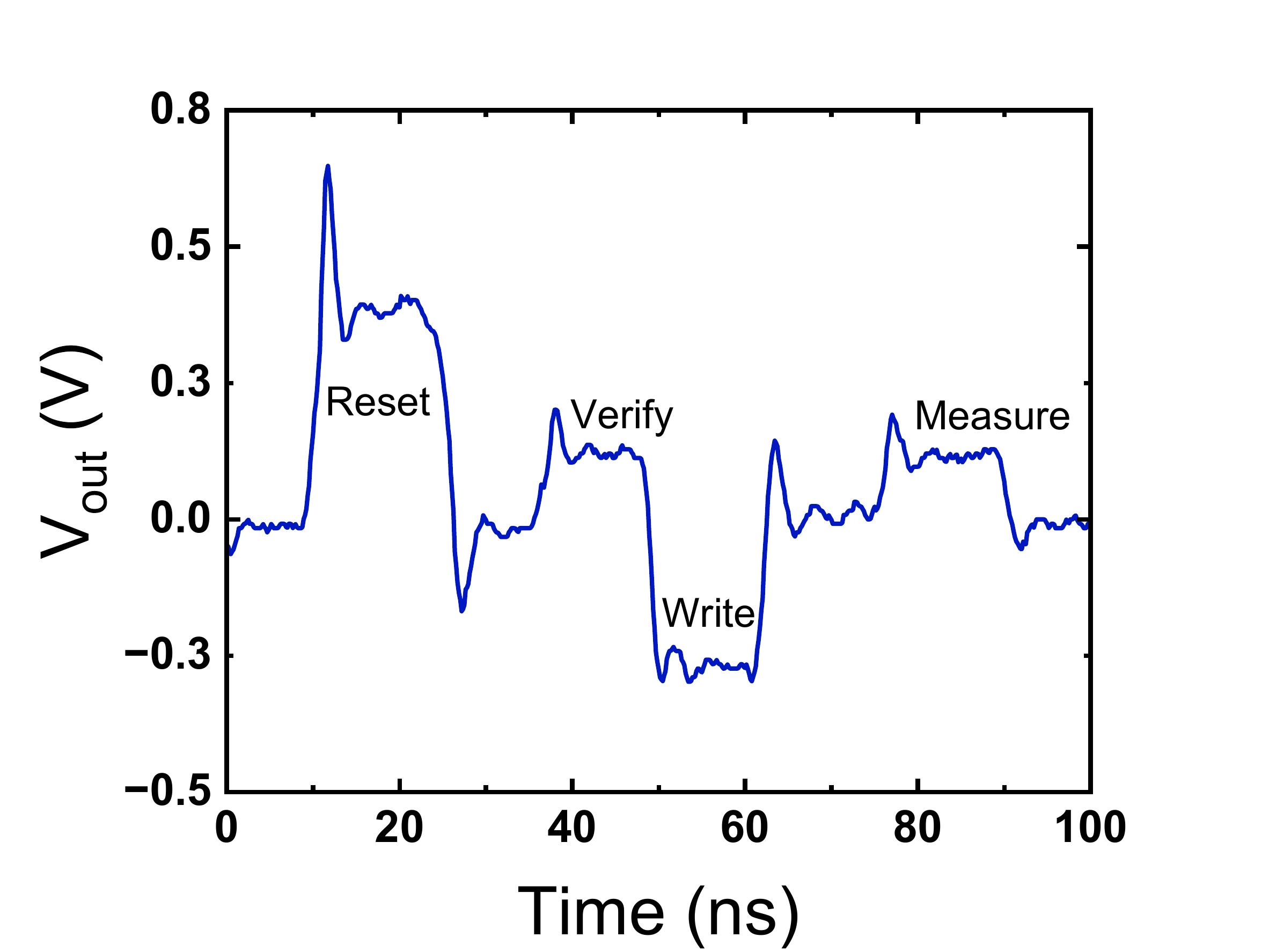} 
    \caption{A single coin flip. This scope trace shows the actuating pulses provided by the custom board at a rate of 10.8MHz.}
    \label{Fig:Fig3}
\end{figure}

Generating a single random bit or one proverbial coin flip consists of the following process:
\begin{enumerate}
    \item Resetting the MTJ to a known state. The AP state here.
    \label{Step1}
    \item Verifying that the MTJ has been reset to the AP state. 
    \label{Step2}
    \item A stochastic write, a short pulse of a specific voltage and duration. 
    \label{Step3}
    \item Measuring the resulting junction state, AP (bit 0) or P (bit 1).
    \label{Step4}
\end{enumerate}
The corresponding output voltage sequence is shown in Fig.~\ref{Fig:Fig3}. The reset step (step~\ref{Step1}) pulse amplitude was set such that we recorded zero reset errors during the verify phase (step~\ref{Step2}), which consists of a lower amplitude output voltage used to determine the junction state. The pulse amplitude for the stochastic write (step ~\ref{Step3}) is set specifically so that the probability of changing states is as close to 50\% as possible. The verify and measure pulses (which are both read pulses) were set high enough to supply a sufficient current to discern the state of the junction but low enough that they do not disturb the junction's state.

To generate one trillion random bits, we had to flip two trillion proverbial coins. The factor of two is necessary to debias the resulting bitstream with an XOR operation, as explained in the next section. The pulse sequence described above --- with a frequency of 10.9MHz --- was applied for 50 hours, resulting in a $2\times 10^{12}$ random bits.

\section{RESULTS AND ANALYSIS}

\begin{figure}[h]
    \centering\includegraphics[width=0.93\linewidth]{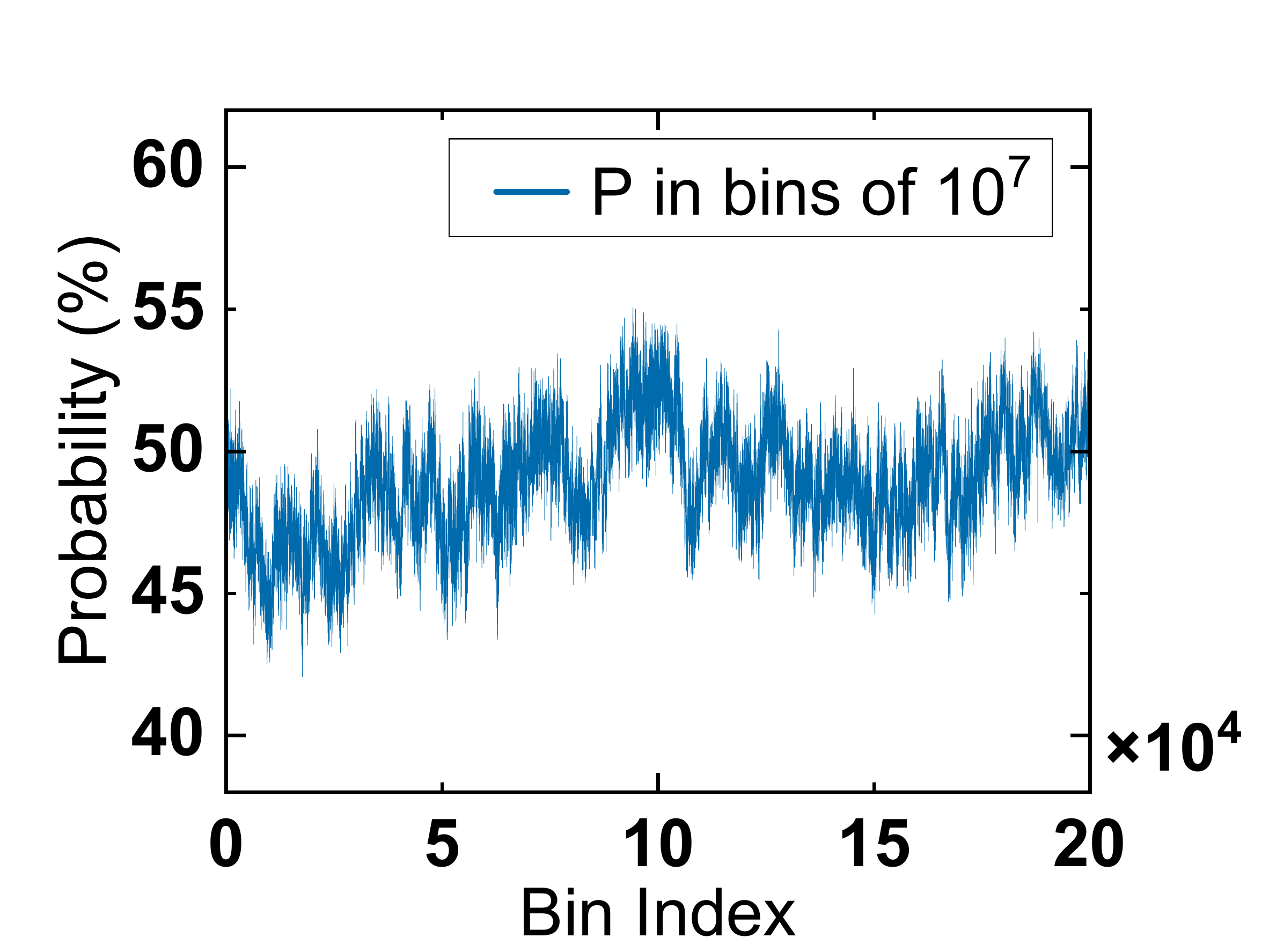} 
    \caption{Probability of switching vs bin index for a total of 2 trillion bits which corresponds to a 50-hour experiment.}
    \label{Fig:Fig4}
\end{figure}

The switching probability for bins of $10^7$ bits was computed and is plotted in Fig.~\ref{Fig:Fig4}. Over the course of the data acquisition, there is a drift from the intended switching probability of 50\%. This drift is greater than expected based on changes in the temperature in the lab during the experiment, as we have measured and characterized the $dP/dT$ in Ref.~~{[}Rehm 2023{]}, and we are still investigating its origin. Nonetheless, these variations can be remedied with one XOR operation.

\begin{figure}[h]
    \centering\hspace*{0.71cm}\includegraphics[width=0.96\linewidth]{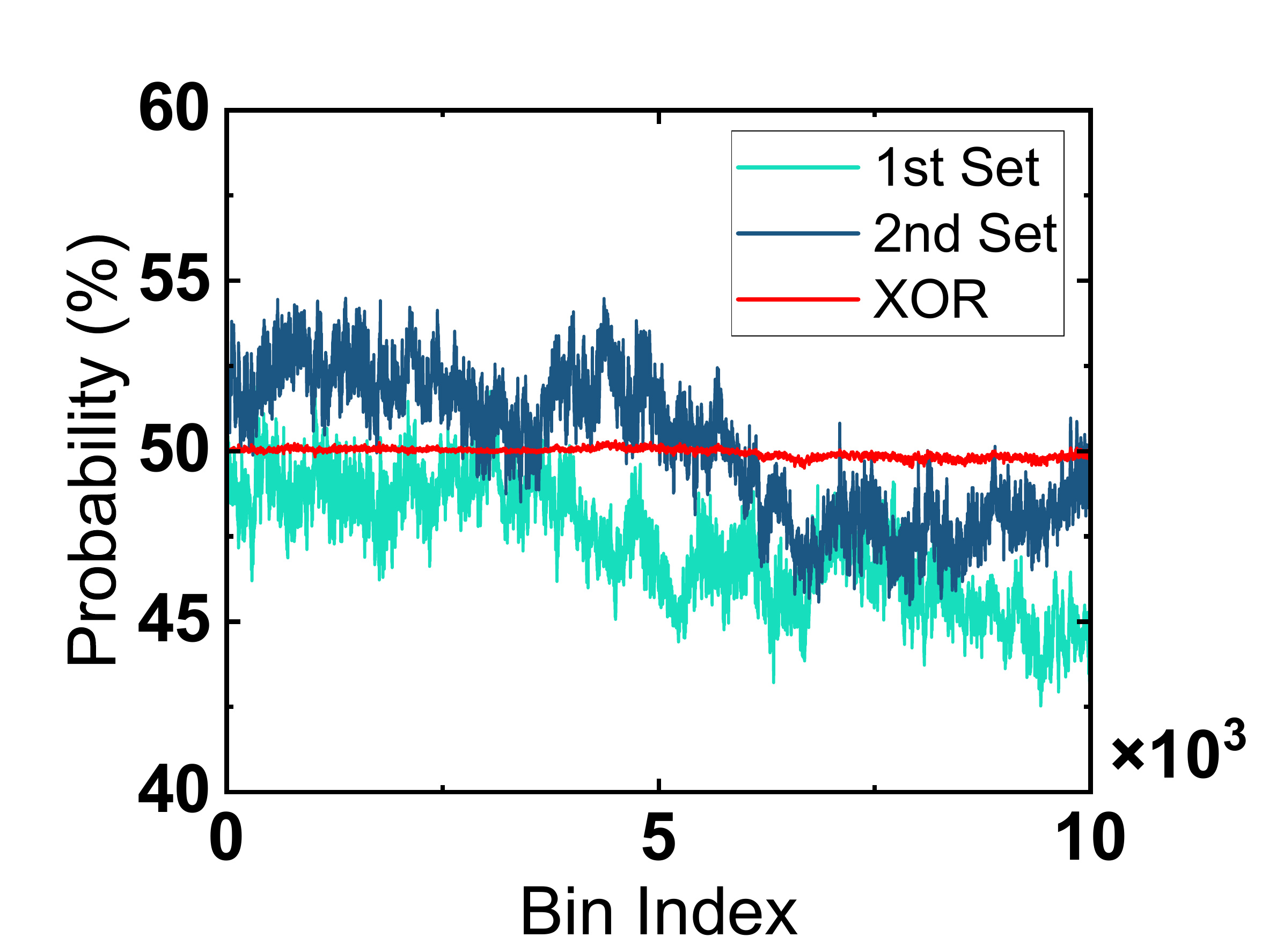} 
    \caption{Switching probability after one XOR operation of the first and second data set of the bitstream.}
    \label{Fig:XOR}
\end{figure}

The XOR operation was performed post-experiment by splitting the bitstream into two halves (the chronological first and second terabits) and XORing them. The results are shown in Fig.~\ref{Fig:XOR} for the first $2\times 10^{11}$ bits in the stream. This operation made the probability of switching significantly closer to 50\% while, of course, reducing the number of total bits by a factor of two. We characterize the probability bias by $\epsilon$, its deviation from 0.5, i.e., $P=0.5+\epsilon$. For the first data set $\epsilon=0.027$ and the second data set $\epsilon=-0.0015$. The XORed data has a significantly reduced bias of $\epsilon=0.0004$.

We tested the resulting bitstreams using the NIST statistical test suite {[}Bassham 2010{]}. This suite encompasses a range of tests, both frequency and non-frequency-related. Frequency-related assessments gauge the occurrence of `ones' within the dataset, while non-frequency tests are designed to identify specific patterns. The results are shown in Table I. The raw data passes only 4 out of 13 tests. In contrast, the XORed data passes all NIST tests. 

\begin{table}
\caption{NIST Statistical Test Suite for randomness results on raw data and XOR'd data. The criteria for passing each test across multiple trials is a 95\% or higher success rate {[}Bassham 2010{]}.}
\footnotesize
\centering
\begin{tabular}{ |p{4.5cm}||p{1.15cm}|p{1.15cm}| }
\hline
\bf Test name / XOR stages & 0 & 1\\
\hline
Frequency (Monobit) & \cellcolor{red}5/100 & \cellcolor{green}98/100 \\
Frequency within a Block & \cellcolor{red}69/100 & \cellcolor{green}100/100 \\
Run & \cellcolor{red}0/100 & \cellcolor{green}100/100 \\
Longest Run of Ones in a Block & \cellcolor{red}22/100 & \cellcolor{green}100/100 \\
Binary Matrix Rank & \cellcolor{green}99/100 & \cellcolor{green}100/100 \\
Discrete Fourier Transform (Spectral) & \cellcolor{green}97/100 & \cellcolor{green}100/100 \\
Non-Overlapping Template Matching & \cellcolor{red}104/148 & \cellcolor{green}142/148 \\
Overlapping Template Matching & \cellcolor{red}67/100 & \cellcolor{green}98/100 \\
Maurer's Universal Statistical  & \cellcolor{red}0/100 & \cellcolor{green}100/100 \\
Linear Complexity & \cellcolor{green}99/100 & \cellcolor{green}99/100 \\
Serial & \cellcolor{green}198/200 & \cellcolor{green}200/200 \\
Approximate Entropy & \cellcolor{red}52/100 & \cellcolor{green}98/100 \\
Cumulative Sums & \cellcolor{red}10/200 & \cellcolor{green}196/200 \\
\hline
\end{tabular}
\end{table}

\section{CONCLUSION}\label{conclusion}
This demonstration that an FPGA with a custom board can efficiently control and sample the state of MTJ devices advances the field and, as we have emphasized, the application of stochastic MTJs. Our FPGA-custom board solution can be used directly in applications that require random numbers, reducing computational loads. For example, in generating and conditioning random bits, the XOR operation could be implemented directly on the firmware of the FPGA to debias the bits in real-time rather than after the fact in the software. Further, several NIST tests could be implemented on the FPGA to verify the bits as they are being produced. There are also possibilities for fast in-situ tuning of device operating parameters to reduce the effect of environmental impacts and device-to-device variations. Finally, the speed of our setup, scale, and low hardware cost will enable a broader range of experimental studies of MTJs, including faster device characterization and, thus, optimization.

\section{ACKNOWLEDGMENTS}
We acknowledge support from the DOE Office of Science (ASCR/BES), Microelectronics Co-Design project COINFLIPS, and the Office of Naval Research (ONR) under Award No. N00014-23-1-2771.

\end{document}